\documentclass{iau} 
\usepackage{graphicx}
\usepackage{natbib,hyperref}

\title[Feeding and feedback in nuclei of galaxies] 
{Feeding and feedback in nuclei of galaxies}

\author[A. Audibert et al.]   
{Anelise Audibert$^1$, Fran\c{c}oise Combes$^2$, Santiago Garc{\'i}a-Burillo$^3$ \and Kalliopi Dasyra$^4$}

\affiliation{$^1$IAASARS, National Observatory of Athens, Penteli, Greece, email:{\tt anelise.audibert@noa.gr} \\[\affilskip]
$^2$Observatoire de Paris, LERMA, Coll{\`e}ge de France, CNRS, PSL University, UPMC, Paris \\ $^3$ Observatorio Astron{\'o}mico Nacional (OAN-IGN)- Observatorio de Madrid, Madrid, Spain\\ $^4$ Department of Astrophysics, Astronomy \& Mechanics, University of Athens, Athens, Greece}

\pubyear{2020}
\volume{359}  
\jname{Galaxy evolution and feedback across different environments}
\editors{T. Storchi-Bergmann, R. Overzier, W. Forman \& R. Riffel, eds.}
\begin{document}

\maketitle

\begin{abstract}
Our aim is to explore the close environment of Active Galactic Nuclei (AGN) and its connection to the host galaxy through the morphology and dynamics of the cold gas inside the central kpc in nearby AGN. 
 We report Atacama Large Millimeter/submillimeter Array (ALMA) observations of AGN feeding and feedback caught in action in NGC613 and NGC1808 at high resolution (few pc), part of the NUclei of GAlaxies (NUGA) project. We detected trailing spirals inside the central 100\,pc, efficiently driving the molecular gas into the SMBH, and molecular outflows driven by the AGN. We present preliminary results of the impact of massive winds induced by radio jets on galaxy evolution, based on observations of radio galaxies from the ALMA Radio-source Catalogue.
\end{abstract}

\firstsection 

\section{Overview: NUGA project}

The key elements in galaxy evolution are  the interplay of the fuelling of SMBH at the center of galaxies and the subsequent feedback from their AGN. Gas inflows into the center of galaxies can fuel the SMBH and the energy input  by the AGN can trigger subsequent feedback. One of the outstanding problems is to identify the mechanism that drives gas from the disk towards the nucleus, removing its large angular momentum (as discussed in \citet{wada04} and \citet{jogee06}, for instance). Feedback processes can be responsible of regulating the SMBH growth \citep{croton06} and explain the  co-evolution of SMBH and their host galaxies, which is now well established by the tight M-$\sigma$ relation \citep{mag98,mcma13}. Recent discoveries of massive molecular outflows \citep[e.g.,][]{fiore17,flu19} have been promoting the idea that winds may be major actors in sweeping the gas out of galaxies, in agreement with theoretical predictions of AGN-driven wind models \citep[see][]{fau12, zub12}. 

Nearby low luminosity AGN (LLAGN) are ideal laboratories to explore the details of outflow and inflowing gas mechanisms. In the NUGA project, we have performed high resolution observations ($\lesssim0.1''$) of the CO(3-2) and dense gas tracers emission with ALMA in a sample of 7 nearby  LLAGN. The sample spans more than a factor of 100 in AGN power (X-ray and radio luminosities), a factor of 10 in star formation rate (SFR), and a wide range of galaxy inner morphology \citep{combes19}. Our goal is to probe feeding and feedback phenomena in these LLAGN, through the study of the morphology and kinematics of the cold molecular gas in galaxy disks and the characterization of the mechanisms driving gas inflows and/or outflows.

We mapped the CO(3-2) and HCN(4-3), HCO$\rm ^+$(4-3) and CS(7-6) emission and compared the morphology of the cold gas to optical images from HST and ionised and warm molecular gas observed in the near-infrared (NIR) with SINFONI. We derived the rotation curves and have modelled the observed velocity field of the CO(3-2) line emission in the galaxy disks in order to find patterns of non-circular motions that could be associated to streaming motions of inflowing gas and/or outflow signatures. To estimate the fuelling efficiency, we have computed the gravitational potential from the stars within the central kpc, from the HST images. Weighting the torques on each pixel by the gas surface density observed in the CO(3-2) line has allowed us to estimate the sense of the angular momentum exchange and its efficiency. In this work, we focus on the study of two individual objects: NGC\,1808 and NGC\,613.

\section{Nuclear trailing spiral in NGC1808}

\begin{figure}
\begin{center}
 \includegraphics[width=0.9\textwidth]{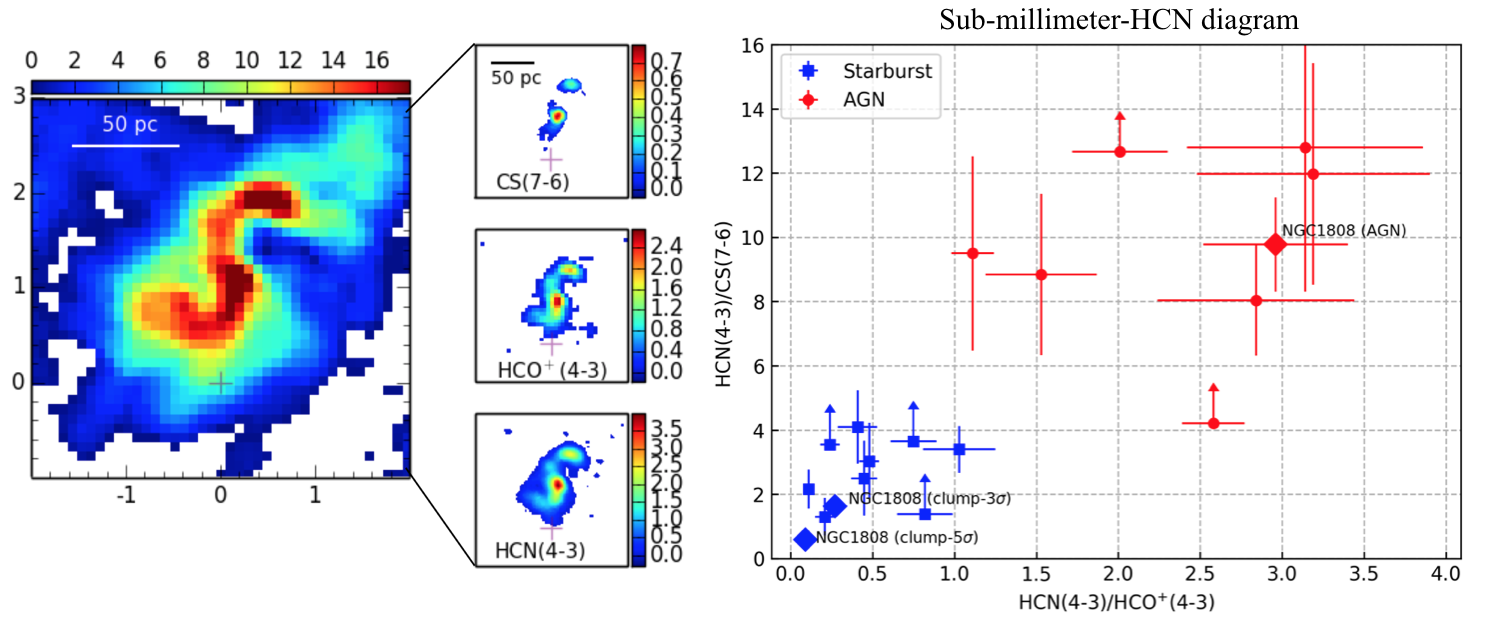} 
 \caption{\textit{Left}: a zoomed $4''\times4''$ region of CO(3-2) intensity map for NGC1808 and dense gas tracers CS(7-6),
HCO$\rm ^+$(4-3) and HCN(4-3). \textit{Right:} submillimetre-HCN diagram \citep{izumi16} using the line intensity ratios $\rm R_{HCN/HCO^+}$ and $\rm R_{HCN/CS}$. We include the line ratios of NGC1808 (diamonds) measured at the centre, or ``AGN", and in a clump detected 140 pc north-west of central position in the dense tracers.}
   \label{sub}
\end{center}
\end{figure}

The ``hot spot'' H\,{\sc ii}/Sy galaxy NGC\,1808 was studied using ALMA Cycle 3 observations at 12\,pc spatial resolution. The CO(3-2) is distributed in a patchy ring at a radius 350\,pc, that is most prominent in the south part and another broken ring at 180\,pc. They are connected by multiple spiral arms. Inside the star-forming ring, a 2-arm spiral structure is clearly detected at $\sim$50\,pc radius (left panel of Fig.\,\ref{sub}), as presented in \cite{ane17}. The nuclear spiral region corresponds to the peak of the velocity dispersions ($\sigma\gtrsim$100\,km/s). The CO morphology shows a remarkable resemblance between the ionised and warm molecular gas along the star forming ring at $\sim4''$ radius, traced by the Pa\,$\alpha$ and H$_2$ emission with SINFONI \citep{busch17}. We found that the nuclear spiral is kinematically decoupled from the larger disk, the position angle being tilted from 323$^\circ$ to close to 270$^\circ$.  

Previous CO(1-0) ALMA observations reported a molecular outflow in the central $\sim$250\,pc \citep{salak16}, but we did not detect outflow signatures in our high-resolution observations. The velocities are mainly due to circular rotation and some perturbations from coplanar streaming motions along the spiral arms.

We confirm the HCN enhancement in circumnuclear molecular gas around AGN, by measuring the HCN(4-3)/HCO$\rm^+$(4-3) and HCN(4-3)/CS(7-6) intensity ratios in the sub-millimetre diagram \citep{izumi16}. We find that the nuclear region of NGC\,1808 presents line ratios that indicate excitation conditions typical of X-ray dominated regions in the vicinity of AGN (Fig.\,\ref{sub}). What is remarkable in our observations, is that the nuclear trailing spiral is even more contrasted in the dense gas tracers. The two-arm spiral structure is also detected in the residual maps in the NIR by \citet{busch17}, supporting the scenario of gas inflow towards the nucleus of NGC\,1808.

\section{Nuclear trailing spiral and molecular outflow in NGC613}

\begin{figure}
\begin{center}
 \includegraphics[width=\textwidth]{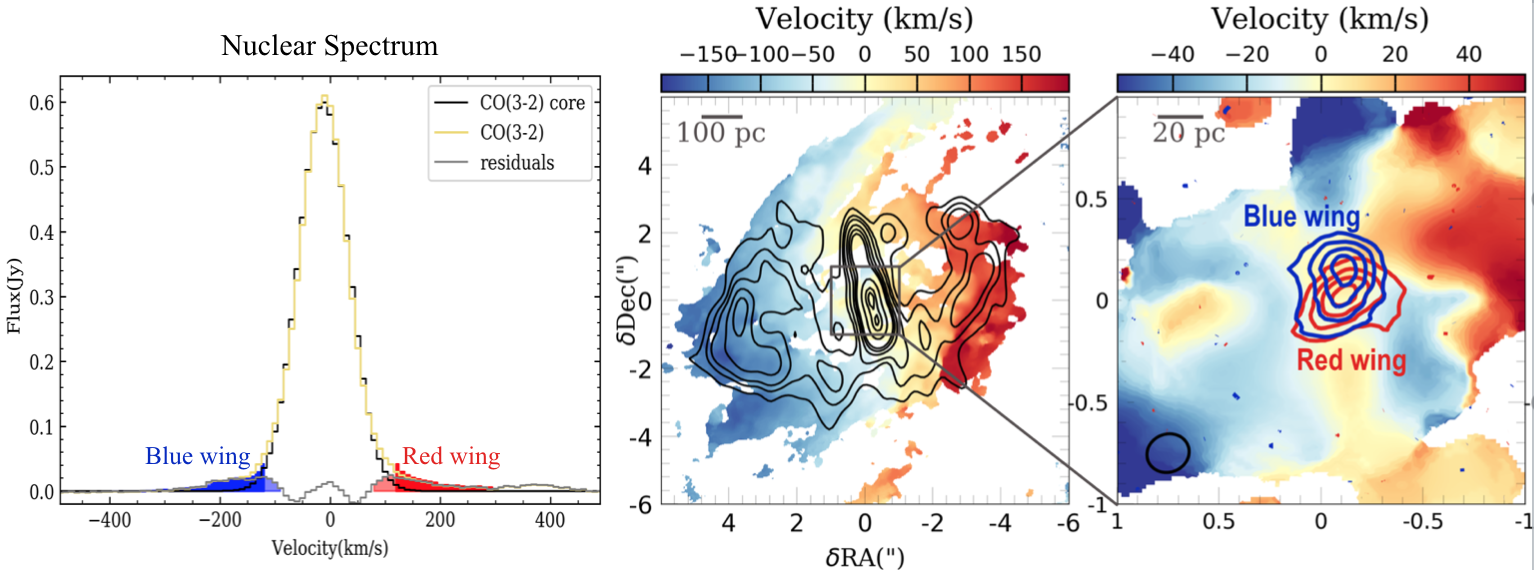} 

 \caption{\textit{Left:} the nuclear CO(3-2) spectrum extracted in a $0.28''$ region. The blue (-400 to-
120 km/s) and red (120 to 300 km/s) wings are associated to the outflow. \textit{Middle and right:} the velocity distribution of the CO(3-2) emission with the VLA radio contours at 4.86 GHz and a $2''\times2''$ zoom of the velocity distribution and the contours of the blue and red wings emission.}
   \label{out}
\end{center}
\end{figure}

In the Seyfert/nuclear starburst galaxy NGC\,613, we have combined ALMA Cycles 3 and 4 observations at a spatial resolution of 17\,pc \citep{ane19}. The morphology of CO(3-2) line emission reveals a 2-arm trailing nuclear spiral at $\rm r\lesssim$100\,pc and a circumnuclear ring at $\sim$350\,pc radius, that is coincident with the star-forming ring seen in the optical images. The molecular gas in the galaxy disk is in a remarkably regular rotation, however, the kinematics in the nuclear region is very skewed.  We find broad wings in the nuclear spectra of CO and dense gas tracers, with velocities reaching up to $\pm$300\,km/s, associated with a molecular outflow emanating from the nucleus ($r\sim$25\,pc, Fig.\,\ref{out}). 

\begin{figure}
\begin{center}
 \includegraphics[width=\textwidth]{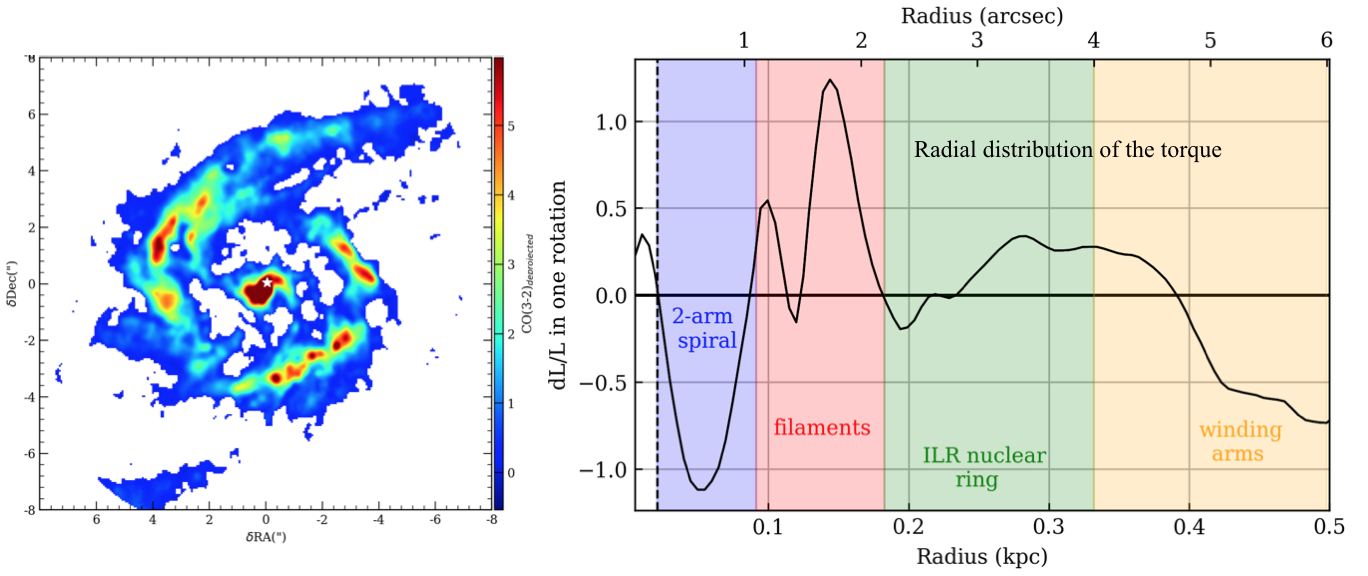} 
 \caption{\textit{Left:} deprojected image of the CO(3-2) emission of NGC\,613 with the morphological features in the \textit{right} panel, showing the radial distribution of the torque, quantified by the fraction
of the angular momentum transferred from the gas in one rotation, $dL/L$.}
   \label{torq}
\end{center}
\end{figure}

We derive a molecular outflow mass $M_{out}$=2$\times$10$^6$M$_\odot$ and a mass outflow rate of  $\dot{M}_{out}=$27\,$\rm M_\odot yr^{-1}$. The molecular outflow energetics exceed the values predicted by AGN feedback models: its kinetic power corresponds to $P_{K,out}=$20\%$L_{AGN}$ and the momentum rate is $\dot{M}_{out}v\sim400L_{AGN}/c$. The outflow is mainly boosted by the AGN through entrainment by the radio jet, but given the weak nuclear activity of NGC\,613, we proposed that we might be witnessing a \textit{fossil outflow}, resulted from a strong past AGN that now has already faded.  
From 25 to 100\,pc, the nuclear trailing spiral observed in CO emission inside the Inner Lindblad Resonance (ILR) ring is efficiently driving gas towards the center. The gravitational torques exerted in the gas show that the gas loses its angular momentum in a rotation period, i.e., in $\sim$10\,Myr dynamical timescale (Fig.\,\ref{torq}). NGC\,613 is a remarkable example of the complexity of fuelling and feedback mechanisms in AGN: given the relative short flow timescale, $\rm t_{flow}\sim10^4$\,yr, the molecular outflow could be a response of the inflowing gas, and eventually acts to self-regulate  the gas accretion. 

\section{Summary}

Among the total NUGA sample of 8 galaxies (including the prototypical Sy\,2 galaxy NGC\,1068 studied by our group), there is evidence of outflows in half of the sample, namely, NGC\,613, NGC\,1433 \citep{combes13}, NGC\,1068 \citep{santi14,santi19} and NGC\,1808 (detected in CO(1-0) by \citet{salak16}, but not confirmed in our high resolution analysis in CO(3-2) emission). The mass outflow rates range from $\sim$1-70\,$\rm M_{\odot} yr^{-1}$, and we confirm the expectations from theoretical models, even in the case of LLAGN: the mass load rates of the outflows increase with the radio power and the AGN luminosity. In the case of the H\,{\sc ii}/Sy\,2 galaxy NGC\,1808, the weakest active object among the detections, the outflow is most likely to be starburst-driven. However, in the other three galaxies, the nuclear SFRs are not able to drive the observed outflows and the properties of the flow require an AGN contribution. Therefore we favour the AGN-driven scenario, in particular the radio mode, where the molecular flow is entrained by the interaction between the radio jet and the interstellar medium (ISM).  At the same time, the observed outflows could  regulate gas accretion in the CND and in short timescales quench the star formation in the nuclear rings, maintaining the balance between gas cooling and heating.

The molecular galaxy disk morphologies reveal the presence of contrasted nuclear rings in the totality of the sample. These rings are quite often the spots of nursery of stars, i.e. usually associated with high SFRs and young star formation, most commonly observed in the optical and NIR. The nuclear rings detected in CO(3-2) emission are usually at the ILR, and in a few cases located at the inner ILR of the nuclear bar, with radius varying from $\sim$170 to $800$\,pc. Since all galaxies in the sample are barred, with different bar strengths, the detection of molecular rings provides evidence of the efficiency of torques due to the bar, driving and piling up the cold gas in rings to eventually form new stars. Although bars are very efficient to drive the gas to a few hundreds of pc scales, an additional mechanism is necessary to bring the gas to the very center and feed the modest black holes at the center of these LLAGN. We find clear evidence of nuclear trailing spirals in 3 galaxies inside the ILR or inner ILR: NGC\,613, NGC\,1808 and NGC\,1566 \citep[presented in][]{combes14}. Previous works have computed the torques in NGC\,1365 and NGC\,1433. In the case of NGC\,1365, it was possible to show that the gas is inflowing to the center, driven by the bar, on a timescale of 300\,Myr \citep{Tabatabaei2013}. For the Sy\,2 galaxy NGC\,1433, the gas is driven towards a nuclear ring of 200 pc radius, at the inner ILR of the nuclear bar, and viscous torques could drive the gas infall towards the very center \citep{combes13,smajic14}.

The project will notably benefit from the improving in the statistics by joining forces with the Galactic Activity, Torus and Outflow Survey (GATOS: \url{gatos.strw.leidenuniv.nl}). GATOS is also mapping the CO(3-2) and HCO$\rm ^+$(4-3) emission with ALMA in the circumnuclear disks of 20 Seyfert galaxies, selected from a ultra-hard X-ray sample, with similar spatial resolution of $0.1''$. Together, NUGA and GATOS will provide a wider range of AGN luminosities and Eddington ratios to explore the connection of inflowing/outflowing gas and molecular tori properties to the host galaxies.

\section{The ALMA Radio-source Catalogue*}

The importance of radio jets in shaping the galaxy evolution have been highlighted in this IAU Symposium. The interaction between radio jets with the ISM has been revealed that relativistic jets can drive molecular and atomic gas outflows, as in the case of the radio bright Seyfert IC\,5063 \citep{mor15,kal16}. ALMA observations have even revealed previously unknown jets thanks to collimated molecular outflows detected in CO \citep[e.g. in NGC 1377 and  ESO 420-G13,][respectively]{aalto16,juan20}. To quantify the impact of radio jets on host galaxies, we built a representative sample of radio galaxies observed with ALMA, the ALMA Radio-source Catalogue, even exploring calibrators. New CO detections, even at high velocities are discovered in this sample.

\begin{scriptsize}
\vspace{2mm}\hspace{-3mm}$^*$This project has received funding from the Hellenic Foundation for Research and Innovation (HFRI) and the General Secretariat for Research and Technology (GSRT), under grant agreement No 1882.
\end{scriptsize}

\end{document}